# Nontraditional hypsometric equation

Hing Ong[1]* and Paul E. Roundy[1]

[1]Department of Atmospheric and Environmental Sciences, University at Albany, State University of New York, Albany, New York

Corresponding author: Hing Ong (hwang2@albany.edu)

**Abstract**

This study corrects the hypsometric equation by restoring the nontraditional terms to relax the hydrostatic approximation. The nontraditional terms include one Coriolis term and two metric terms in the vertical momentum equation. The hypsometric equations with and without the nontraditional correction are used to calculate the geopotential height of pressure levels using more than 300,000 selected tropical rawinsonde profiles. With westerlies between two pressure levels, the thickness of the layer increases, which reduces the upward pressure gradient forces to balance the upward nontraditional Coriolis forces; the opposite is true for easterlies. Hence, zonal winds are negatively correlated with traditional geopotential height biases aloft. At 500 hPa, for example, traditional geopotential height error in the tropical troposphere is on the order of at least 0.5 m, which is considerable with respect to geopotential height variability in tropical large-scale flow, on the order of 10 m to 15 m.

## 1. Introduction

The hypsometric equation is important for interpreting radiosonde observations, which are traditionally amongst the most reliable upper atmospheric data sources. This equation relates geopotential height difference between two pressure levels to mean virtual temperature in between, using the hydrostatic approximation (e.g., Holton and Hakim, 2013). In radiosonde observations, the measured pressure, temperature, and humidity can be used to calculate the geopotential height of pressure levels via the hypsometric equation. This method is regarded as a reliable estimation of geopotential height, as the geopotential height records in many rawinsonde databases are derived using the hypsometric equation by default, and direct estimates based on the global positioning system (GPS) are only occasionally used. Furthermore, since many of the current global atmospheric models use the hydrostatic approximation, the model-analyzed geopotential height conforms to hypsometric-derived height.

However, the significance of the nontraditional Coriolis terms (NCTs) in tropical large-scale flow affects the validity of the hydrostatic approximation (Ong and Roundy, 2019, hereafter

OR19). NCTs occur because of the use of a frame of reference that rotates with Earth; they turn eastward motion upward and turn upward motion westward, and vice versa. Using this frame of reference, the centrifugal acceleration is conventionally included in the gravity term. Moreover, using a frame of reference that rotates with air motion, the centrifugal acceleration can be corrected by the vertical component of NCTs along with metric terms. OR19 proposed an indirect measure to validate the hydrostatic approximation. First, the ratio of the NCT to the traditional Coriolis term in the zonal momentum equation is on the order of 10% in tropical large-scale dynamics (OR19; see also Hayashi and Itoh, 2012; White and Bromley, 1995), which encourages the inclusion of the NCT in the zonal momentum equation into models, relaxing the traditional approximation (Eckart, 1960). Then, to conserve energy, the relaxation of the traditional approximation requires the inclusion of the NCT in the hydrostatic equation, which then becomes quasi-hydrostatic (White and Bromley, 1995). Consequently, the hydrostatic approximation is indirectly associated with a zonal wind bias in tropical large-scale flow on the order of 10% (OR19).

Furthermore, the NCT in the vertical momentum equation may directly affect the weight of flowing air mass. Eötvös' (1919) experiments showed that an object weighs more when moving westward and weighs less when moving eastward because of the NCT, so-called Eötvös effect. Eötvös noticed this effect in a gravity dataset measured on ships moving around the globe, and then conducted an experiment measuring gravity on two ships in the Black Sea moving westward and eastward, which confirmed the theory. According to Persson's (2005) review, papers from 1894 to 1923 discussed meteorological applications of the Eötvös effect. The discussion ceased after Szolnoki (1923) concluded that the Eötvös effect is far smaller than the *total* weight of flowing air mass. However, the *spatial variability* of weight of flowing air mass can be a better reference when atmospheric dynamics are concerned, because the deviation of geopotential height from a spatially averaged hydrostatic state drives the horizontal flow.

OR19 motivates further assessment of the Eötvös effect on geopotential height data derived from rawinsonde observations using the hypsometric equation. Westerly wind yields upward NCT, which is balanced by an extra downward pressure gradient force (PGF', where the prime denotes perturbation due to the inclusion of NCT) using the quasi-hydrostatic approximation. This reduces the total upward PGF, so thickness between two pressure levels increases. The opposite is true for easterly winds. The aim of this study is to measure the significance of error in hypsometric-derived height due to omission of the Eötvös effect in tropical large-scale dynamics, using the spatial variability of geopotential height as a reference.

The rest of this paper is organized as follows. Section 2 derives a nontraditional hypsometric equation and analyzes the scale of the NCT in it. Section 3 describes data and methods used in this study. Section 4 compares the geopotential height results using the nontraditional and traditional hypsometric equation and discusses the limits of this study. Section 5 presents conclusions.

**2. Derivation and Scale Analysis**

The nontraditional hypsometric equation is derived from the quasi-hydrostatic equation (White and Bromley, 1995);

$$\frac{1}{\rho}\frac{\partial p}{\partial z} + g(1 + A) = 0, \tag{1a}$$

$$A = -\frac{1}{g}\left(2\Omega u \cos\vartheta + \frac{u^2+v^2}{r}\right), \tag{1b}$$

where the variables are defined as follows: $\rho$, density; $p$, pressure; $z$, geopotential height; $r$, distance from Earth center; $g$, gravity acceleration; $\Omega$, Earth rotation rate; $\vartheta$, latitude; $u$, zonal velocity; $v$, meridional velocity. $A$ denotes the nontraditional correction due to the NCT, $2\Omega u \cos\vartheta$, and the metric terms, $\frac{u^2+v^2}{r}$; the metric terms are small but included for dynamical consistency. On the other hand, the hydrostatic approximation omits the nontraditional correction, i.e., $A = 0$. This study focuses on vertical profiles, so $p$ and $z$ are one-to-one functions.

Equation (1a) conforms to the hydrostatic equation, so the derivation of the nontraditional thickness equation is straightforward following the traditional derivation (e.g., Holton and Hakim, 2013). Apply $\rho = \frac{p}{RT_v}$ ($R$ and $T_v$ denote gas constant for dry air and virtual temperature) and integrate equation (1a) from lower level 1 to higher level 2;

$$z(p_2) - z(p_1) = \int_{p_2}^{p_1} \frac{RT_v}{g(1+A)} d\ln p. \tag{2}$$

Equation (2) is the nontraditional hypsometric equation when the lower bound of the integral is at the surface. For application to rawinsonde profiles, equation (2) is discretized;

$$z(p_2) - z(p_1) = \frac{R\overline{T_v}}{g(1+\tilde{A})} \ln\left(\frac{p_1}{p_2}\right), \tag{3a}$$

$$\tilde{A} = -\frac{1}{g}\left(2\Omega \bar{u} \cos\vartheta + \frac{\bar{u}^2+\bar{v}^2}{r}\right), \tag{3b}$$

where the overbars denote the average of the two levels.

The scale of the nontraditional correction, $A$, may be analyzed as follows: $A \sim 10^{-4}$ in the tropics and midlatitudes, given $g \sim 10$ m s$^{-2}$, $2\Omega \cos\vartheta \sim 10^{-4}$ s$^{-1}$ (valid from the equator to $\sim 60°$ latitude), and $u \sim 10$ m s$^{-1}$. Moreover, given the geopotential height in the troposphere $\sim 10^4$ m, the geopotential height error due to omitting $A$ in these regions should be $\sim 1$ m. Given a horizontal length scale of $\sim 10^6$ m, the error of $\sim 1$ m may be negligible in the midlatitudes, where the large-scale geopotential height variability is $\sim 100$ m, but may be considerable in the tropics, where the variability is $\sim 10$ m; see chapter 18 of Vallis (2017) for a scale analysis, and see Sakaeda and Roundy (2016) for composite height anomaly data in phases of the Madden–Julian oscillation (MJO). The small scale of tropical geopotential height variability is consistent with the corresponding small scale of temperature variability. The tropical temperature variability associated with, e.g., the MJO (e.g., Kiladis et al., 2005) or convectively coupled equatorial waves (CCEWs, e.g., Kiladis et al., 2009) is $\sim 0.3$ K, which is $\sim 10^{-3}$ of the tropospheric temperature. Therefore, this study focuses on the geopotential height error due to omitting $A$ in the tropics. Furthermore, the other NCT, which turns upward motion westward and downward motion eastward, may also play a role in tropical large-scale dynamics. However, such effects cannot be analyzed with hypsometric equations using individual rawinsonde profiles and are beyond the scope of this study.

## 3. Data and Methods

To measure the significance of the nontraditional correction, rawinsonde data from Integrated Global Radiosonde Archive (IGRA) version 2 (Durre et al., 2018) are used. IGRA consists of individual radiosonde or pilot balloon profiles, monthly mean profiles, and sounding-derived parameters. IGRA includes data from more than 2,700 stations, and the earliest data date back to 1905. About 1,000 stations currently report data, and IGRA version 2 is updated in almost real time. IGRA includes data at the standard pressure levels (1000, 925, 850, 700, 500, 400, 300, 200, 150, and 100 hPa are included in this study) and other pressure levels ununiformly distributed between the standard levels. This study uses every available level for upward integration of equation (3a) from the surface but outputs results only at the standard levels. The traditional geopotential height biases are inferred from the traditional results using $A = 0$ minus the nontraditional results using equation (1b).

In this study, only the pressure, temperature, relative humidity, wind direction, and wind speed in the individual rawinsonde profiles from selected stations are used. The first criterion of the station selection is that the latitude must lie between 15°S and 15°N because this study focuses in the tropics. The second criterion of the selection is that the station must be selected for Radiosonde Atmospheric Temperature Products for Assessing Climate (RATPAC, Lanzante et al., 2003) because the selected stations have continual long records. Fifteen stations in IGRA satisfy these criteria. A sampling problem is that the selected stations do not cover a broad range of longitude over the Pacific Ocean. However, more than 300,000 rawinsonde profiles are selected in this study, so a broad range of weather conditions can be covered. Table 1 lists information about the selected stations including identification code (ID), station name, latitude (LAT), longitude (LON), elevation (EL), first (FST) and last (LST) year of rawinsonde record, and number of rawinsonde profiles (NUM) when we accessed the data on 24 June 2019.

**Table 1.** List of selected IGRA stations

| ID | NAME | LAT (°) | LON (°) | EL (m) | FST | LST | NUM |
|---|---|---|---|---|---|---|---|
| ASM00094120 | Darwin Airport | −12.4239 | 130.8925 | 31.4 | 1950 | 2019 | 27900 |
| BPM00091517 | Honiara | −9.4167 | 159.9667 | 55.0 | 1959 | 2011 | 7645 |
| BRM00082332 | Manaus | −3.1500 | −59.9833 | 84.0 | 1967 | 2019 | 17420 |
| COM00080222 | Bogota/Eldorado | 4.7000 | −74.1500 | 2547.0 | 1960 | 2019 | 21756 |
| FMM00091334 | Truk/Caroline Is. | 7.4500 | 151.8333 | 3.0 | 1951 | 2019 | 38002 |
| IOM00061967 | Diego Garcia | −7.3000 | 72.4000 | 3.0 | 1967 | 2006 | 10877 |
| IVM00065578 | Abidjan | 5.2500 | −3.9333 | 7.0 | 1957 | 2019 | 19705 |
| KEM00063741 | Dagoretti Corner | −1.3036 | 36.7597 | 1798.3 | 1957 | 2018 | 18541 |
| NGM00061052 | Niamey-Aero | 13.4833 | 2.1667 | 223.0 | 1953 | 2019 | 28742 |
| PSM00091408 | Koror | 7.3687 | 134.5412 | 53.1 | 1951 | 2019 | 36731 |
| RMM00091376 | Majuro/Marshall Is. | 7.0683 | 171.2942 | 3.9 | 1952 | 2019 | 34949 |
| SGM00061641 | Dakar/Yoff | 14.7300 | −17.5000 | 24.5 | 1949 | 2018 | 32702 |
| SHM00061902 | Wide Awake Field | −7.9667 | −14.4000 | 79.0 | 1946 | 2010 | 17023 |
| SNM00048698 | Singapore/Changi | 1.3667 | 103.9833 | 5.0 | 1955 | 2019 | 31421 |
| THM00048455 | Bangkok Metropolis | 13.7330 | 100.5670 | 4.0 | 1953 | 2008 | 24344 |

To measure the large-scale variability of geopotential height and zonal wind, 6-hourly reanalysis data from ERA-Interim Project (Dee et al., 2011) are used. The 27 available pressure levels from 1000 hPa to 100 hPa are used. A horizontal range of [180°W, 180°E; 15°S, 15°N] and a temporal range from 1979 to 2018 are covered. The horizontal grids are Gaussian with spacings of ~ 0.703°. The Fourier transform is used to filter the reanalysis data for zonal wavenumbers 1 through 10. The large-scale geopotential height variability is measured using the horizontal standard deviation of the filtered data. On the other hand, the large-scale zonal wind variability is measured using the unfiltered average plus and minus the filtered horizontal standard deviation. The measures are different because the averaged geopotential height is dynamically insignificant, but the averaged zonal wind can contribute to the traditional geopotential height biases.

In the above-mentioned methods, the traditional geopotential height biases and the large-scale geopotential height variability are derived from different data. To derive the large-scale geopotential height variability, spatially continuous data are required for the Fourier transform, so reanalysis data are used. To derive the traditional geopotential height biases, reliable surface pressure data are required as the lower bounds for the upward integration, so rawinsonde data are used. Despite lacking a reliable level to start the integration, reanalysis data can be used to depict the horizontal distribution of contributing factors to the integral in equation (2), e.g., the vertical NCT and buoyancy ($-\frac{\rho'}{\rho_0} g$, where the subscript 0 and the prime denote horizontal average and deviation from the average). As a side product of this study, the vertical NCT and buoyancy are regressed upon MJO-filtered precipitation averaged from 15°S to 15°N. GPCP Version 1.3 One-Degree Daily Precipitation Data Set (Mesoscale Atmospheric Processes Branch and Earth System Science Interdisciplinary Center, 2018) is used. The filter band covers zonal wavenumber from 1 to 10 and time period from 30 days to 96 days. The statistical significance is tested at 95% confidence level with two-tailed Student's t-test, in which the equivalent degrees of freedom account for autocorrelation of 1-day lag.

## 4. Results and Discussion

According to equation (3), positive $\bar{u}$ contributes to negative $\tilde{A}$, which further contributes to a larger height difference. Thus, the results of every output level show a negative correlation between the traditional geopotential height biases and the zonal wind average below the level. Because the contribution of zonal wind to geopotential height accumulates during upward integration, the sample standard deviation of the traditional geopotential height biases increases with height; it is 0.39 m at 500 hPa, 0.96 m at 200 hPa, and 1.52 m at 100 hPa. Even a zonal wind at a single level can serve as an indicator of the traditional geopotential height bias aloft. For example, the correlation coefficient between the zonal winds at 700 hPa (U700) and the traditional geopotential height biases at 500 hPa (Z500 biases) is –0.90. The probability density of data points of these two variables is depicted in Figure 1. For most of the data points, U700 varies between ± 30 m s$^{-1}$, and Z500 biases vary between ± 2 m. Most of the data points lie around the linear regression line, and few outliers are present, which can be explained by vertically localized zonal wind maxima. If the zonal wind maximum lies at 700 hPa, U700 will be strong, but Z500 biases will be weaker than the prediction by regression. If the zonal wind maximum lies below 500 hPa but not at 700 hPa, U700 will be weak, but Z500 biases will be stronger than predicted by the regression.

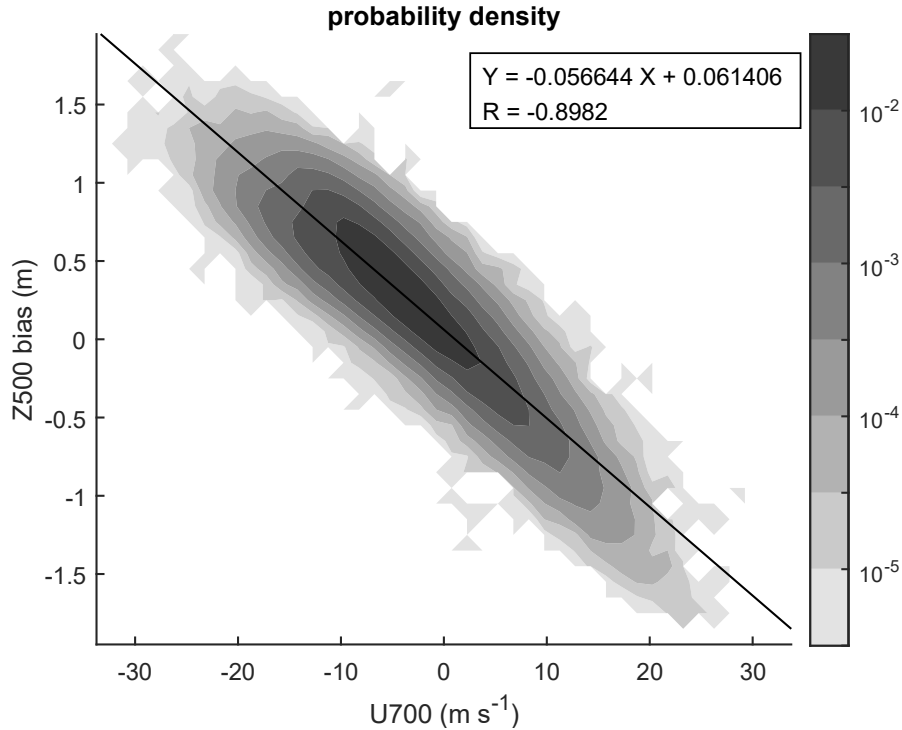

**Figure 1**. Probability density of data points of the traditional geopotential height biases at 500 hPa (Z500 bias) and the zonal winds at 700 hPa (U700). More than 300,000 points are used. The solid line denotes the linear regression line.

The horizontal standard deviation of the large-scale filtered geopotential height (hereafter, filtered variability) at 500 hPa is on average 12.7 m with temporal standard deviation of 2.6 m, and such an order of magnitude is consistent with previous studies (e.g., Sakaeda and Roundy, 2016; Vallis, 2017). Within one standard deviation, the large-scale filtered zonal wind at 700 hPa varies within [–7.9, 1.5] m s$^{-1}$, and such a range corresponds to traditional geopotential height biases at 500 hPa ranging from –0.02 m to 0.51 m, according to the regression line in Figure 1. Such regressed biases are 3% ~ 5% of the filtered variability. However, the regression model is based on rawinsonde data; with the surface as a reliable reference level, zonal wind contributes only upward to traditional geopotential height biases. Without a reliable reference level, zonal wind can contribute both upward and downward, causing the regression method to overestimate the contribution from zonal wind below the pressure level of interest and to ignore the contribution from above. We speculate that the regressed biases at 500 hPa underestimate the biases due to omitting the NCT because they do not account for contributions from zonal winds aloft. Because zonal wind directions are often opposite in tropical upper and lower troposphere, the traditional thickness biases are also opposite. A positive thickness bias below corresponds to a positive geopotential height bias, which may be enhanced by a negative thickness bias above. The opposite is true for a negative thickness bias below. In summary, in this study, the traditional geopotential height biases and the large-scale geopotential height variability are accurate in different contexts (whether a reference level is present), and the contextual difference limits the accuracy of the comparison between them. The next paragraph attempts to exemplify the contextual difference.

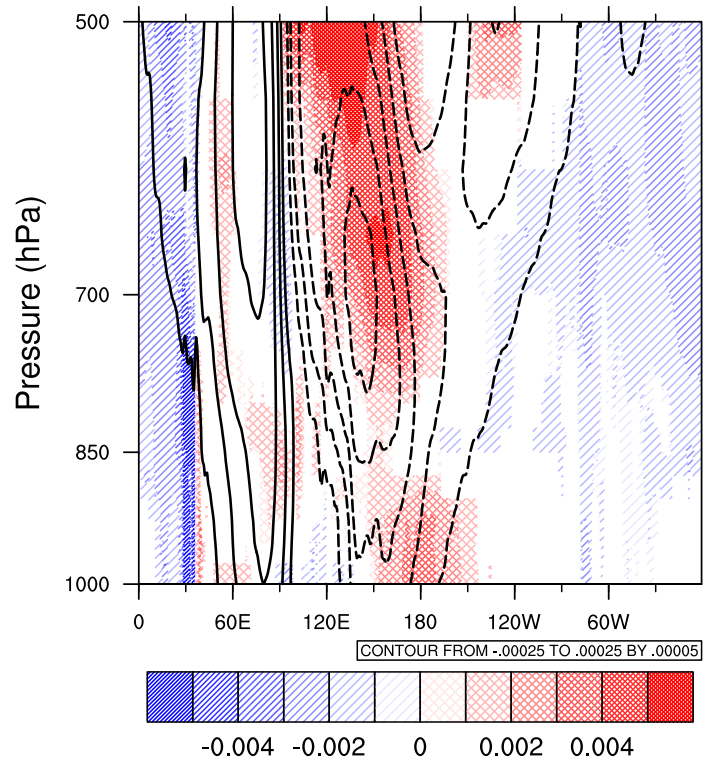

**Figure 2.** Zonal vertical distributions at the equator of the vertical nontraditional Coriolis term (NCT, contours, m s$^{-2}$) and buoyancy (hatching, m s$^{-2}$) regressed upon MJO-filtered tropical precipitation at 90°E. The prediction at one standard deviation of the filtered precipitation is shown. All shown results are significant at 95% confidence level. The solid and dashed contours denote positive and negative values. The zero contour is omitted.

Figure 2 depicts zonal vertical distributions in the lower troposphere at the equator of the vertical NCT and buoyancy regressed upon MJO-filtered tropical precipitation at 90°E. In terms of the absolute value of the maxima (exclude topography-related ones), the vertical NCT is ~ 5% of the buoyancy. Positive NCT dominates in the rear flank (roughly from 45°E to 90°E), where buoyancy signals are sparse and transition from negative to the west to positive to the east. A positively buoyant region is located in the forward flank (roughly from 90°E to 180°E), and the NCT offsets a small fraction of upward buoyancy in this region. This analysis is related to the main topic of this study because the vertical NCT and buoyancy are contributing factors to the thickness between two pressure levels. However, without a reference level, the factors can affect the geopotential height above or below, and to which direction they contribute remains undetermined.

## 5. Conclusions

A previous study suggested that the hydrostatic approximation, which omits the nontraditional Coriolis terms (NCTs), is indirectly associated with a zonal wind bias in model-simulated tropical large-scale flow on the order of 10% (Ong and Roundy, 2019). The present study further assesses a direct effect of the NCT, so-called Eötvös effect, on the geopotential height derived from rawinsonde observations using the hypsometric equation. The corrected hypsometric equation is derived from the quasi-hydrostatic equation, which restores the nontraditional terms to relax the hydrostatic approximation. The nontraditional terms include the NCT and two metric terms in the vertical momentum equation. Both the traditional and nontraditional hypsometric equations are used to calculate the geopotential height of pressure levels using more than 300,000 selected tropical rawinsonde profiles. The traditional geopotential height biases are inferred from the traditional results minus the nontraditional results.

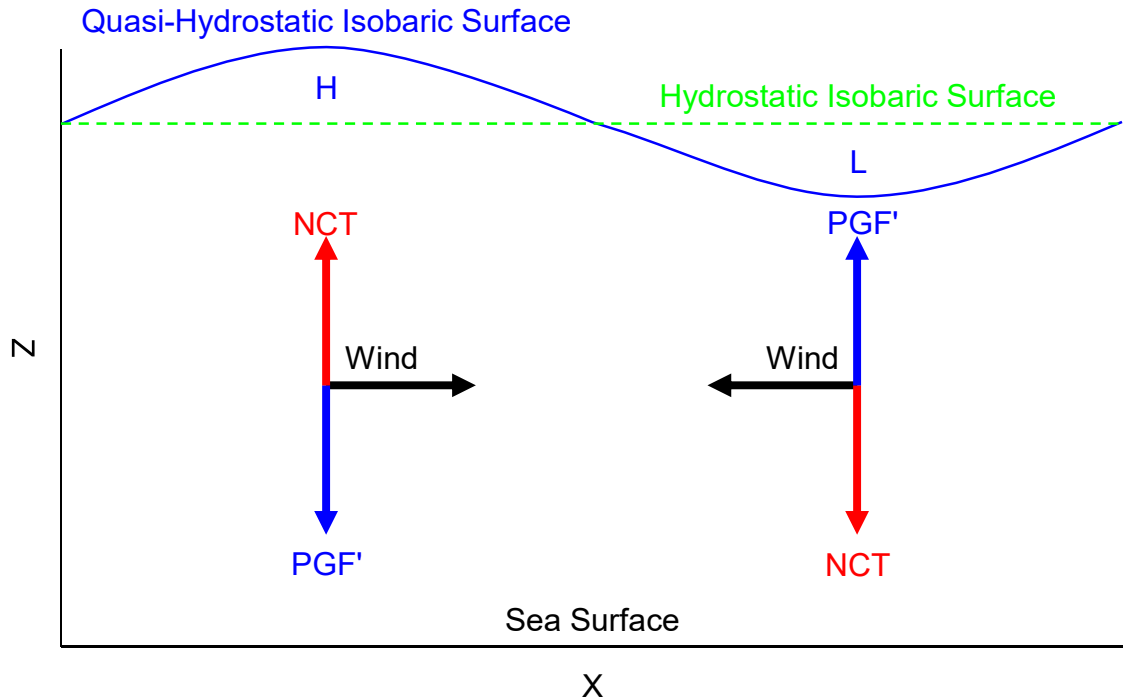

**Figure 3.** A schematic diagram illustrating the effect of the nontraditional Coriolis term (NCT) on the geopotential height of an isobaric surface. See Conclusions for discussion.

The diagram in Figure 3 illustrates the direct effect of the NCT on the geopotential height of an isobaric (constant pressure) surface. Given the sea surface pressure and height, the geopotential height of an isobaric surface aloft can be calculated using either the traditional (hydrostatic) or nontraditional (quasi-hydrostatic) hypsometric equation. For simplicity, consider horizontally homogeneous density. First, assuming hydrostatic balance, the isobaric surface would be flat. Then, consider the quasi-hydrostatic effect of zonal winds between the sea surface and the isobaric surface. A westerly wind yields upward NCT, which is balanced by an extra downward pressure gradient force (PGF'). The downward PGF' corresponds to a high-pressure perturbation (H) aloft, so the quasi-hydrostatic isobaric surface is higher than the hydrostatic one. The opposite is true for an easterly wind. This reasoning can be applied only if a reliable reference level is present to start vertical integration; for example, pressure and height at the surface in rawinsonde data are accurately observed.

The results suggested that zonal winds at 700 hPa can serve as an indicator of the traditional geopotential height biases at 500 hPa because they are negatively correlated. At 500 hPa, for example, traditional geopotential height error in the tropical troposphere is on the order of at least 0.5 m, which is considerable with respect to geopotential height variability on the order of 10 m to 15 m in tropical large-scale flow, including the Madden–Julian oscillation (MJO) and convectively coupled equatorial waves (CCEWs).

**Acknowledgements**

This work was funded by National Science Foundation (grants AGS1757342, AGS1358214 and AGS1128779). IGRA version 2 and ERA-Interim Project data, supporting the conclusions, can be obtained via https://doi.org/10.7289/V5X63K0Q and https://doi.org/10.5065/D6CR5RD9. We thank ECMWF for granting access to ERA-Interim Project data via NCAR Research Data Archive. We thank Hungjui Yu and three anonymous reviewers for discussions and comments.